\documentclass[%
reprint,
superscriptaddress,
 amsmath,amssymb,
 aps,
pra,
]{revtex4-2}

\usepackage{graphicx}
\usepackage{dcolumn}
\usepackage{bm}

\usepackage{dsfont}

\usepackage[colorlinks, linkcolor=red, anchorcolor=blue,urlcolor=blue, citecolor=blue]{hyperref}

\newcommand{\beq}{\begin{equation}}
\newcommand{\eeq}{\end{equation}}
\newcommand{\beqa}{\begin{eqnarray}}
\newcommand{\eeqa}{\end{eqnarray}}
\def\ra{\rangle}
\def\la{\langle}

\begin{document}

\preprint{APS/123-QED}

\title{ Scattering of relativistic electrons and analogies with optical phenomena: A study of longitudinal and transverse shifts at step potentials}

\author{Yue Ban}
\email{ybanxc@gmail.com}
\affiliation{TECNALIA, Basque Research and Technology Alliance (BRTA), 48160 Derio, Spain}
 
\author{Xi Chen}
\email{xi.chen@ehu.eus}
\affiliation{Department of Physical Chemistry, University of the Basque Country UPV/EHU, Apartado 644, 48080 Bilbao, Spain}
\affiliation{EHU Quantum Center, University of the Basque Country UPV/EHU, Barrio Sarriena, s/n, 48940 Leioa, Spain}

\date{\today}

\begin{abstract}
We investigate the behavior of relativistic electrons encountering a potential step through analogies with optical phenomena.
By accounting for the conservation of Dirac current, we elucidate that the Goos-Hänchen  shift can be understood as a combination of two components: one arising from the current entering the transmission region and the other originating from the interference between the incident and reflected beams. This result has been proven to be consistent with findings obtained utilizing the stationary phase method. Moreover, we explore the transverse Imbert-Fedorov shift, by applying both current conservation and total angular momentum conservation, revealing intriguing parallel to the spin Hall effect. Beyond enriching our comprehension of fundamental quantum phenomena, our findings have potential applications for designing and characterizing devices using Dirac and topological materials.

\end{abstract}

\maketitle


\section{Introduction}

The Goos-Hänchen (GH) effect, first described by Isaac Newton \cite{Newton} in his pioneering work on optics and later named after Goos and Hänchen in 1947 \cite{GH1,GH2}, stands as a notable optical phenomenon. Newton's observations laid the groundwork for understanding this effect, which was further substantiated by the theoretical and experimental contributions of Goos and Hänchen. Their studies revealed that when a light beam undergoes total internal reflection at the interface between two different refractive media, it undergoes a slight lateral shift from the position predicted by geometrical optics. Initially explained within the framework of classical wave optics, the GH effect revealed the intricate behavior of light at interfaces \cite{Renard:64,Lotsch:68}.
Another remarkable phenomenon, known as the Imbert-Fedorov (IF) effect, was independently discovered by C. Imbert \cite{imbert1968calculation} and V. Fedorov \cite{rabinovich1955isotopic}. The IF effect pertains to the transverse shift experienced by a totally reflected beam in the plane perpendicular to the incidence plane. This shift arises due to a polarization-dependent phase shift between the electric and magnetic field components of the reflected beam \cite{Schilling,ImbertPRD}. Unlike the GH effect, the IF effect shares similarities with the spin Hall effect of light, both of which emerge from the complex interplay between the polarization properties and the reflection behavior of light \cite{NeilPRD,BeauregardPRD,Hallprl04}. These two optical phenomena, the GH and the IF effects, have been explored in diverse optical systems and materials \cite{Pillon:05,Aiello:08,WenPRA12,ZubairyPRA,Bliokh_2013,Toppel2013,Goswami:16,Ornigotti:18,Wenpra19,Kong:19,JIAO2019239}, including plasmonic structures, photonic crystals, and metamaterials. The understanding and control of these effects have facilitated significant advancements in the field of nanophotonics, optical communications, and sensing technologies.

Following the idea of changing the perspective, the exploration of analogies between quantum mechanics and classical optics has provided a fruitful avenue for understanding and analyzing quantum systems \cite{dragoman2004quantum}. It has allowed us to leverage insights and techniques from optics to gain new perspectives on quantum phenomena, driving advancements in our knowledge and applications of quantum mechanics \cite{DRAGOMAN1999131}. 
Within the framework of quantum mechanics,  the behavior of scattering particles, including Dirac particles and neutrons, has attracted significant attention. Researchers have actively sought to transplant concepts originally observed in classical optics, such as the GH effect \cite{Renard:64,Carter:71,Neutrons} and the IF effect \cite{MillerPRL,FradkinPRD1,FradkinPRD2}, into the context of quantum particle scattering.

With the advent of mesoscopic transport in condensed matter physics, the study of GH and IF effects has expanded to encompass various quantum systems, including semiconductors, graphene, and topological insulators, as reviewed in \cite{Chen_2013,yu2019anomalous}. In the realm of semiconductors, the GH effect for ballistic electrons has been integrated into the analysis of crucial processes such as reflection, transmission, and tunneling, playing a vital role in the design and optimization of electronic devices such as transistors, diodes, and integrated circuits \cite{Wilson}. Furthermore, the GH shifts can be dynamically tuned by external magnetic or electric fields, allowing for the manipulation of electron beams and control of electron transport in semiconductor-based devices \cite{Chenprb08,Chenprab11}. Similarly, the GH shifts and their tunability have been investigated in graphene-based systems, offering a wide range of applications \cite{Beenakkerprl,Kaiprl,Yelinprb}. Remarkably, the GH shfits and IF shifts as well become particularly interesting in the presence of crossed Andreev reflection at a metal/superconductor interface when combined with graphene \cite{Shenyuanprb18a,Shenyuanprb18b}. In recent times, their connection with phenomena such as anomalous Hall effect, Berry curvature \cite{Nomuraprb}, screw scattering, and side jump \cite{vignale2010ten,skewscatteringprl}, has garnered considerable attention in Weyl semimetals \cite{Shengyuanprl20,Xieprl15,Yangprl15,Yaoprb19,Jianprb17,Dongreprb22}. 
These phenomena are closely linked to the unique electronic and topological properties, thus leading to potential applications in spintronics and topological phenomena. 
Therefore, gaining fundamental insight into their physical mechanics from different perspectives is essential.

In this article, we present a comprehensive investigation into the GH and IF shifts exhibited by relativistic electrons as they scatter by step potentials. The motivation to focus on relativistic electrons, govern by the Dirac equation, stems from the intriguing possibility of unveiling fruitful phenomena resulting from the unique coupling of spin and orbital angular momentum \cite{Bliokhprl11,Bliokhprl12,Zofiaprl17,Bliokhpra17,Barnettprl,Alexanderpra20}, particularly the complex dynamics at the intersection of relativistic quantum mechanics and the distinctive properties of electron vortices.
Building upon the foundation of the Dirac equation, we analyze the scattering of relativistic electrons. Due to the absence of spin flip in the transmission through the potential barrier \cite{de2009planar}, our study delves into two-dimensional Dirac diffusion at a single scattering interface. 
Exploiting the explicit spin-dependent amplitudes and phases of reflection and transmission, we address the analytical aspects of the GH shift in term of conservation of Dirac current, and compare the results obtained from the stationary phase method (SPM) \cite{Artmann}. Also we connect the IF shift obtained from the current conservation to spin Hall effect, in terms of spin-to-orbital angular momentum conversion. Our results will not only provide the physical insight of the GH and IF shifts,  and the potential applications of these shifts in Dirac materials and topological systems, which share a fundamental similarity with Dirac particles.

\section{Preliminary results}

\begin{figure}[t]
\begin{center}
\scalebox{0.18}[0.18]{\includegraphics{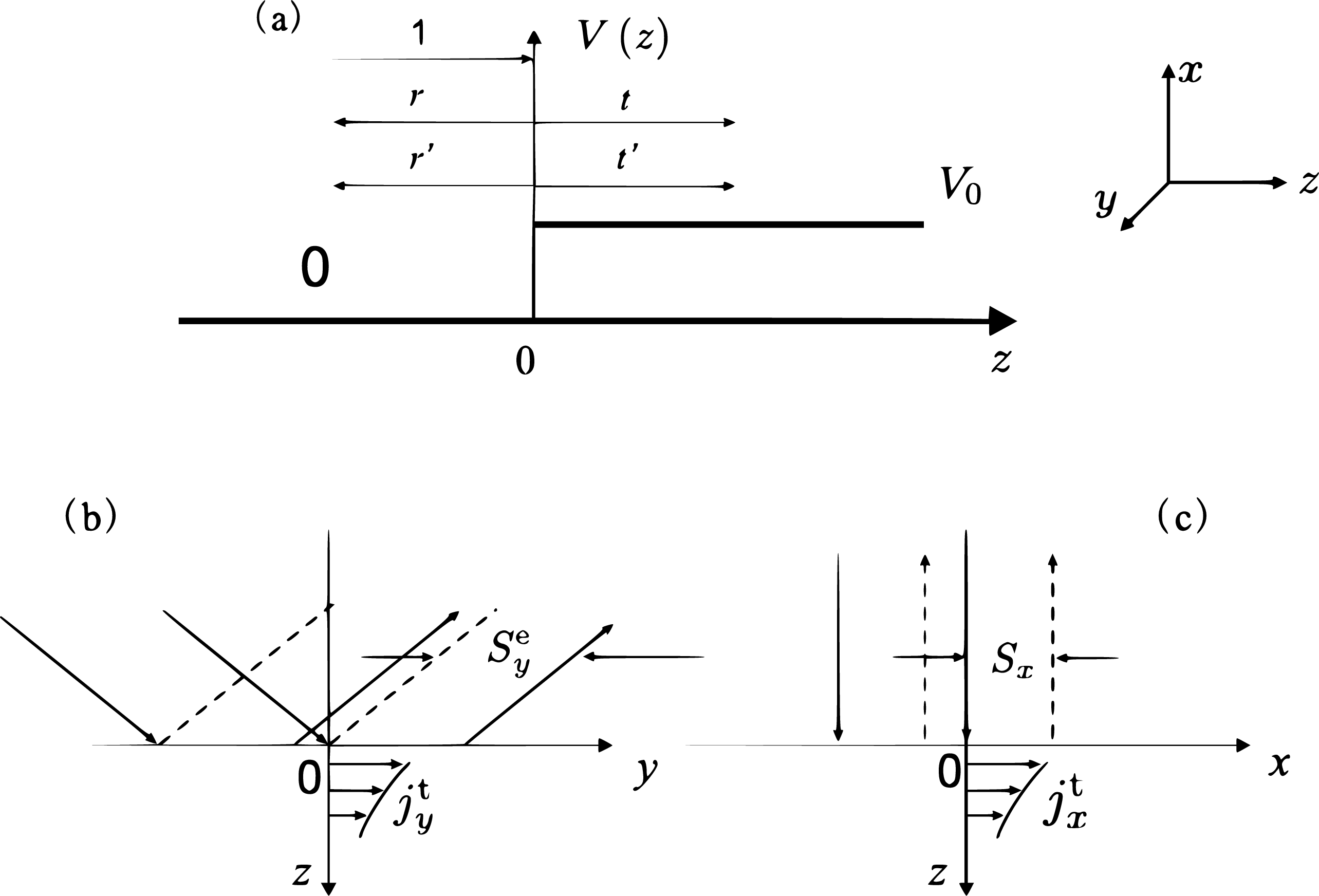}}
\caption{(a) Schematic diagram of relativistic electrons encountering a step potential, depicting spin-flip reflection and transmission. Projection in the $yoz$ plane (b) and in the $xoz$ plane (c), showcasing the GH ($S_y$) and IF ($S_x$) shifts, respectively, along with the corresponding currents. }
\label{potential}
\end{center}
\end{figure}

Our objective is to investigate the longitudinal GH shift and the transverse IF shift of relativistic electrons as they encounter a step potential, as illustrated in Fig. \ref{potential}. The step potential is defined by
\begin{eqnarray} \ V\left(z\right)= \left\{
\begin{array}{ll}
0  &~~ (z<0){,}
\\
\nonumber
V_0 &~~ (z>0){.}
\end{array}
\right.
\end{eqnarray}
where $V_0 > 0$. In general, the relativistic electron satisfies the stationary Dirac equation, $H \psi =E \psi$, with the Hamilton \cite{greiner2000relativistic,thaller2013dirac},
\begin{equation}
H= \vec{\alpha} \cdot \vec{p} + \beta m^2,  
\end{equation}
where $m$ is the mass, $E$ is the energy, and with a convenient choice based on the Pauli spin matrices,
\beqa
\alpha_{i} = \left(\begin{array}{cc}
0 &  \sigma_{i}
\\
\sigma_{i} & 0
\end{array} \right), ~~
\beta = \left(\begin{array}{cc}
I & 0
\\
0 & -I
\end{array} \right).
\eeqa
Here $\vec{\alpha}$ presents the Dirac matrices $\alpha_{i}$ ($i=x,y,z$) corresponding to the $z$, $y$ $z$ directions with Pauli matrices $\sigma_i$, respectively, and $I$ is identity matrix. For simplicity, we choose natural units with $\hbar = c = 1$ and the time-dependent factor $e^{-i E t}$ is suppressed.
We consider this investigation for two-dimensional scattering arbitrary with an incidence angle $\theta$, taking into account the central plane wave of an relativistic electron beam with finite width.
The wave function of the plane wave under the consideration along the $z$ direction is assumed to be
\beqa
\label{psizin}
\psi_{\textrm{in}}(z) = \left(
\begin{array}{cccc}
1
\\
0
\\
 \frac{p_{z}}{E+m}
\\
 \frac{i p_{y}}{E+m}
\end{array}
\right) e^{i p_{z} z},
\eeqa
where $p_x=0$, $p_z$ and $p_y$ are momentum components along the $z$ and $y$ directions, respectively. This wave function is the free particle solutions to the Dirac equation, i.e. \cite{thaller2013dirac}
\begin{equation}
-i \alpha_z \psi'(z) + \alpha_y p_y \psi(z) + \beta m \psi(z) = E \psi(z).
\end{equation}
The incident plane wave has the polarization in the $z$ direction, denoted as spinor ``$|\uparrow \rangle$", up to an overall normalization factor $ 1/\sqrt{\mathcal{N}}$ (see appendix \ref{appendixA}). The dispersion relation is given by $E = \sqrt{p_y^2 + p_z^2 + m^2}$, and the incidence angle satisfies $\tan\theta = p_y/p_z$. Upon encountering the step potential, the reflected plane wave consists of a coexistence of the spin-up and spin-down components, as shown in Fig. \ref{potential}(a). Similarly, 
the reflected wave can be expressed as
\beqa
\label{psizre}
\psi_{\textrm{r}}(z) = 
r\left(
\begin{array}{cccc}
1
\\
0
\\
\frac{-p_{z}}{E+m}
\\
\frac{i p_{y}}{E+m}
\end{array}
\right)  e^{-i p_{z} z}
+r'\left(
\begin{array}{cccc}
0
\\
1
\\
\frac{-i p_{y}}{E+m}
\\
\frac{p_{z}}{E+m}
\end{array}
\right) e^{-i p_{z} z},
\eeqa
where $r$ and $r'$ represent the amplitudes of the spin-up $|\uparrow\rangle$ and spin-down $|\downarrow \rangle$ components, respectively.
\begin{figure}[t]
\begin{center}
\scalebox{0.85}[0.85]{\includegraphics{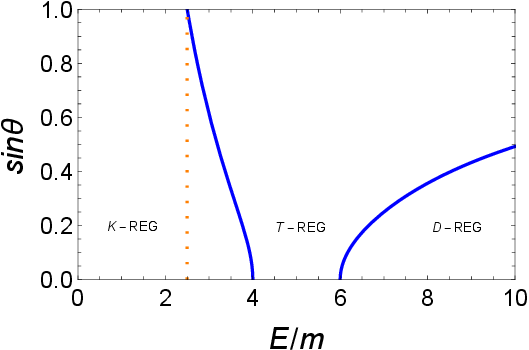}}
\caption{Separation of the various energy zones by fixing respectively the incident angle $\theta$ and the potential $V_0$, where $V_0=5 m$. Two blue solid lines are the critical angles of the existence of tunneling, and orange dotted line presents $E=2.5m$, corresponding to $\theta_c =\pi/2$, solved from Eq. (\ref{criticalangle}). }
\label{E-sintheta}
\end{center}
\end{figure}
In the region $z > 0$, the wave function in the $z$ component can be written as
\beqa
\label{psiztr}
\psi_{\textrm{t}}(z) = t\left(
\begin{array}{cccc}
1
\\
0
\\
\frac{q_{z}}{E-V_0+m}
\\
\frac{i q_{y}}{E-V_0+m}
\end{array}
\right) e^{i q_{z} z} 
+t'\left(
\begin{array}{cccc}
0
\\
1
\\
\frac{-i q_{y}}{E-V_0+m}
\\
\frac{-q_{z}}{E-V_0+m}
\end{array}
\right)  e^{i q_{z} z},~~~
\eeqa
where $t$ and $t'$ represent the amplitudes of the spin-up $|\uparrow\rangle$ and spin-down $|\downarrow \rangle$ components, respectively. Here, $q_y = p_y$, and $q_z = \sqrt{(E-V_0)^2 - q_y^2 - m^2}$ in the region of potential. 

To facilitate the analysis, we introduce an effective mass $m^* = \sqrt{p_y^2 + m^2}$. The energy zones can then be divided into three regions:
\beqa
\left\{ \begin{array}{lll}
&\textit{D}:& E>V_0+m^*,
\\
&\textit{T}:& V_0-m^*<E<V_0+m^*,
\\
&\textit{K}:& V_0-m^*>E.
\end{array}
\right.
\eeqa
as depicted in Fig. \ref{E-sintheta}, the diffusion, tunneling, and Klein regions \cite{greiner2000relativistic}, respectively, which is determined based on the relations between $E$ and $\sin\theta$ with a fixed value of $V_0 = 5m$.

The boundaries between the tunneling region and the other two regions determine the conditions for generating evanescent waves in the transmission region. From an optical analogy, since the $y$-component of the wave vector is continuous, the critical angle for total reflection can be defined as
\begin{equation}
\label{criticalangle}
\theta_c = \arcsin\left(\sqrt{\frac{(E-V_0)^2-m^2}{E^2-m^2}}\right).
\end{equation}
The critical angle depends on the incident energy and coincides with the boundaries of the tunneling region. For example, Eq. (\ref{criticalangle}) determines the critical angles for total reflection $\theta_c =0.408638 \simeq 23.4^{\circ}$ with the given parameters $E=8.5 m$, $V=5m$. In the diffusion and Klein cases oscillatory solutions exist everywhere,
thus we concentrate on the tunneling case for studying GH and IF shifts, characterized by real exponential solutions in the potential region.

Considering the continuity of the wave function at the interface, e.g., $\psi_{\textrm{in}}(0^-) + \psi_{\textrm{r}}(0^-)= \psi_{\textrm{t}}(0^+)$, the scattering coefficients can be derived as follows:
\begin{align}
\label{r}
r &= \frac{(p_y^2 + m^2 + mE) V_0}{(E+m)(p_z^2+p_z q_z - V_0 E)}, \\
t &= \frac{p_z^2(E-V_0+m) + p_z q_z (E+m)}{(E+m)(p_z^2+p_z q_z - V_0 E)}, \\
\label{r'}
r' &= \frac{i p_y p_z V_0}{(E+m)(p_z^2+p_z q_z - V_0 E)}, \\
t' &= r',
\end{align}
where $q_z = i \kappa$ for $\theta>\theta_c$. In the diffusion region, $\theta < \theta_c$, the conservation of current along the $z$ direction leads to the following relation among the scattering coefficients:
\begin{equation}
|r|^2 + |r'|^2 + \frac{q_z (E+m)}{p_z (E-V_0+m)}(|t|^2+|t'|^2) = 1.
\end{equation}
For $\theta > \theta_c$, the transmitted waves exhibit an evanescent behavior, which corresponds to the tunneling region where total reflection occurs, characterized by 
\begin{equation*}
|r|^2+|r'|^2 = 1.
\end{equation*}
Indeed, the intriguing phenomenon arises when relativistic electrons encounter a potential step with oblique incidence, where $q_y \neq 0$. This scenario, as evidenced by the coefficient $r'$ in Eqs. (\ref{r'}), highlights the crucial role played by spin flip in the presence of $q_y \neq 0$. The spin flip effect stands out as a distinct feature of relativistic Dirac particles experiencing oblique incidence, which profoundly impacts their scattering behavior, particularly the IF shift as discussed later.

\section{longitudinal GH shift}
\label{GH}

In the subsequent section, we will utilize the energy flux method \cite{Renard:64, yasumoto1983new} to provide a comprehensive explanation of these unique phenomena from the perspective of Dirac current analysis.
Drawing on the energy flux model in optical fashion \cite{yasumoto1983new,Fedoseyev:86,Seshadri:88,Chen:12}, we discern that the longitudinal GH shift can be attributed to two distinct components. The first component originates from the current within the evanescent wave present in the transmission region, while the second component arises from the interference between the incident and reflected beams. To begin, we will perform calculations to determine the GH shift resulted from the current within the evanescent wave.

\begin{figure}[t]
\begin{center}
\scalebox{0.85}[0.85]{\includegraphics{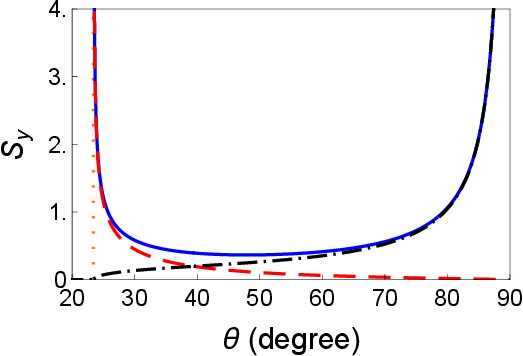}}
\caption{Dependence of the longitudinal GH shift $S_y=S_y^\textrm{e}+S_y^{\textrm{ir}}$ on the incidence angle $\theta$, provided by $E_0=8.5 m$, $V_0=5 m$ and $m=1$ for simplicity, where the shifts contributed from  $S_y^\textrm{e}$ (dashed, red), and  $S_y^{\textrm{ir}}$ (dot-dashed, black) are compared to the one obtained from stationary phase method $S_y^p = S_y$ (solid, blue). Orange dotted line denotes the GH shift around critical angle $\theta_c$ for the eye.}
\label{Sgh}
\end{center}
\end{figure}

In general,  the Dirac current is defined as $\vec{j}= \psi^{\dag} \vec{\alpha} \psi$, where $\psi$ is the wave function and $\vec{\alpha}$ is the Dirac matrix as described above.
To calculate the GH shift resulting from the current within the evanescent wave ($\theta> \theta_c$), we can first use Renard's model \cite{Renard:64}. As a result, the GH shift is given by
\begin{equation}
S_y^\textrm{e} = \frac{P}{j^{\textrm{r}}_z} = \frac{\int_0^{\infty}j_y^{\textrm{t}} dz}{j^{\textrm{r}}_z}.
\end{equation}
Here as shown in Fig. \ref{potential}(b), $P$ is the total current inside the transmission region in the $y$ direction, $j_y^{\textrm{t}} = \psi^\dag_{t} \alpha_y \psi_{t}$ is the $y$ component of the $z$-dependent current in the evanescent field, and $j^{\textrm{r}}_z = \psi^\dag_{\textrm{r}} \alpha_z \psi_{\textrm{r}}$ is the $z$ component of the reflected current. It is assumed that the incident beam has finite width in the incidence plane. When total reflection occurs, the current  enters the transmission part, experiences displacement, and returns to the $z<0$ region. Due to this, the shift of the reflected beam calculated by Renard's energy flux method \cite{Renard:64} is given by
\begin{equation}
S_y^\textrm{e} =\frac{(E+m)}{2(E-V_0+m)}\left(\frac{|t|^2+|t'|^2}{\kappa}-\frac{t^*t'+t't^*}{p_y} \right),
\end{equation}
Additionally, the current presented here along the $y$ direction is also attributed to the interference between the incident beam and the reflected beam,
\beqa
\label{Jir}
j_{\textrm{ir}} &=& \frac{1}{2} \textrm{Re} \left[\psi^\dag_{\textrm{in}} \alpha_y \psi_{\textrm{r}} + \psi^\dag_{\textrm{r}} \alpha_y \psi_{\textrm{in}}\right].
\eeqa
We can integrate $j_{\textrm{ir}}$ over $z$ from some distance point $-l$ to obtain the total particle flux:
\begin{eqnarray}
P_{\textrm{ir}} &=& \int_{-l}^0 j_{\textrm{ir}} dz, 
\end{eqnarray}
where $j_{\textrm{ir}}$ can be directly calculated from Eq. (\ref{Jir}), 
{\small 
$$   
j_{\textrm{ir}} = \frac{4}{E+m}\left[p_y |r| \cos(2p_{z}z-\varphi_r) - p_{z} |r'| \sin(2p_{z}z-\varphi_{r'})\right],
$$
}
with $\varphi_{r}$ and $\varphi_{r'}$ being the phase of coefficients in reflection for different spinors, see Eqs. (\ref{r}) and (\ref{r'}). 
As $P_{\textrm{ir}}$ contains trigonometric terms, to obtain the average flux and eliminate rapid oscillation, we divide it by the length scale of order $\lambda_z = 2\pi /p_{z}$:
\begin{equation}
\la P_{\textrm{ir}} \ra = \frac{\int_{0}^{\lambda_z} P_{\textrm{ir}} dl  }{\lambda_z} =  \frac{2 |r'|}{E+m} \cos \varphi_{r'}-\frac{2p_{y}|r|}{p_{z}(E+m)}\sin\varphi_r,
\end{equation}
Therefore, the shift part induced by the interference is given by
\begin{equation}
S_y^{\textrm{ir}} = \frac{\la P_{\textrm{ir}} \ra}{j^{\textrm{in}}_x} = -\frac{p_{y}}{p_{z}}\textrm{Im}[r] + \frac{1}{p_{z}}\textrm{Re}[r'].
\end{equation}
From the perspective that the shift comes from the current along the interface, the whole shift due to total reflection is $S_y = S_y^\textrm{e}+S_y^\textrm{ir}$. As a matter of fact, one can also calculate the GH shift from the point $y=0$ by using the SPM \cite{Artmann}, which is defined as
$S^{p}_y = - \partial \varphi/ \partial p_y$ \cite{Artmann},  with $\varphi$ being 
$ \varphi_{r}$ and $ \varphi_{r'}$, the phase shifts of reflected electrons with spin-up and spin-down polarizations, respectively. From Eqs. (\ref{r}) and (\ref{r'}), we have
\begin{equation}
\varphi_{r} = \tan^{-1} \left( \frac{\kappa p_z}{p^2_z - V_0 E}\right), ~~ \varphi_{r} = \varphi_{r'}+\pi/2,
\end{equation}
thus these two GH shifts are identical, yielding
\begin{equation}
\label{S}
S_y^p \!=\frac{(E^2-m^2) (V_0 E- p_{z}^2)\sin2\theta} {2 \kappa [(p_{z}^2-V_0 E)^2+p_{z}^2 \kappa^2]} \!- \frac{\kappa (p_{z}  p_{y} + V_0 E\tan \theta)}{(p_{z}^2-V_0 E)^2+p_{z}^2 \kappa^2}.
\end{equation}
It is evident from the longitudinal GH shift derived from the SPM, $S^p_y$, see Eq. (\ref{S}), is exactly the same as $S_y$ from the current conservation, as illustrated in Fig. \ref{Sgh}. When the incidence angle approaches $\theta_c$, the self-interference shift vanishes, while near grazing incidence it plays a dominant role.

\section{transverse IF shift}
\label{IF}

Next, we shall turn to the characteristics of the reflected beam in the $yox$ plane, by considering transverse IF shift. Similar to the previous analysis in the $yoz$ plane, with some width in the $yox$ plane, the reflected beam undergoes a displacement caused by the current in the transmission region along the $x$ direction, given that $\theta>\theta_c$. This transverse IF shift is expressed as
\beqa
\label{Sif}
S_x= \frac{\int_0^\infty j_x^\textrm{t}dz}{j_z^\textrm{r}} = \frac{i (E+m)}{2 p_y (E-V_0+m)}(t^* t'-t'^* t),
\eeqa
where $j_x^\textrm{t} = \psi_\textrm{t}^\dag \alpha_x \psi_\textrm{t}$ represents the Dirac current along the $x$ direction originating from the transmission region, as indicated in Fig. \ref{potential}(c).

\begin{figure}[t]
\begin{center}
\scalebox{0.85}[0.85]{\includegraphics{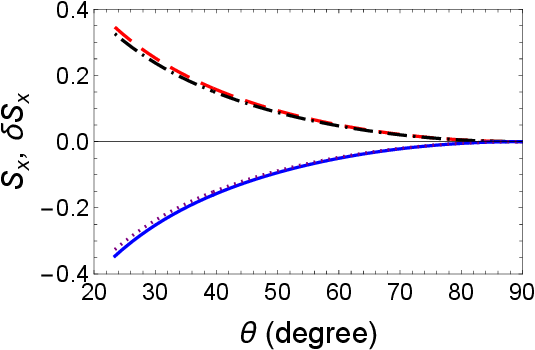}}
\caption{Dependence of the transverse IF shift $S_x$ on the incidence angle $\theta$, where 
blue solid and red dashed lines correspond to spin-up and spin-down polarizations of incidence. For comparison, the shifts $\delta S_x$ calculated from Eq. (\ref{spinHall}) are also denoted by purple dotted and black dotted-dash line with two different polarizations, respectively. All parameters are the same as those in Fig. \ref{Sgh}.}
\label{SIF}
\end{center}
\end{figure}

As the reflected beam is parallel to the incident beam in the $yox$ plane, the interference between the reflected beam and the incident beam does not contribute to the IF shift. Also, it is worth noting that the IF shift is much smaller than the GH shift in the tunneling case. The IF shift reaches its maximum but remains a finite value for an incidence angle very near the critical angle in the evanescent case. Moreover, the IF shift occurs for incidences below the critical angle of incidence and even at larger angles, which is not depicted here. In contrast, the GH shift goes to infinity as the incidence angle approaches the critical angle. The incident beams, with opposite polarizations represented by the spinors ``$|\uparrow \rangle$" and ``$|\downarrow \rangle$", have corresponding eigenvalues $\sigma_z=\pm 1$, respectively.
As illustrated in Fig. \ref{SIF}, the IF shift, as derived from Eq. (\ref{Sif}), exhibits the same magnitude but opposite direction for different polarizations. 
This intriguing behavior bears resemblance to the IF shift observed in the context of right- and left-circularly polarized light beams during total reflection \cite{FradkinPRD1,FradkinPRD2,ImbertPRD}.
Moreover, in our case, the initial polarization can be described as the eigenstates of helicity, e.g.,
$\vec{\sigma} \cdot \vec{p} = \pm 1$, which is the superposition $\cos(\theta/2) |\uparrow \rangle \pm i \sin (\theta/2) |\downarrow \rangle$. As a result, the IF shifts, corresponding to the eigenstates of helicity, become $S'_x= S_x \cos \theta$.
This further implies the possibility of controlling the IF shifts by varying the polarization of the incident beam.

Besides, the IF shift can be derived by considering the conservation of total angular momentum (TAM). In principle, the total angular momentum operator for the relativistic Dirac particle can be expressed as $\vec{J} = \vec{L} + \vec{\Sigma}$,
where $\vec{L}$ is the orbital angular momentum operator, and $\vec{\Sigma}$ is the spin operator. 
In our scenario, the $z$ component of the TAM, $J_z = L_z + \Sigma_z$, is conserved, i.e., 
$\langle J_z^\textrm{in} \rangle = \langle J^\textrm{r}_z \rangle $. Since $p_x=0$, we have $L_z = x p_y$ and obtain, see appendix (\ref{appendixA}),
\begin{equation}
\label{spinHall}
\delta S_x = \langle x^\textrm{r}_0 \rangle - \langle x^\textrm{in}_0 \rangle = \frac{\mathcal{N}}{2 p_y} \left[1-(|r|^2- |r'|^2) \right].
\end{equation}
Noting that if the angle of incidence is $\theta$, the angle of the reflected particle $\pi- \theta$ is then used here. As shown in Appendix \ref{appendixA}, we have used unitary Foldy-Wouthuysen (FW) transformation to separate the positive and negative energy sub-spaces to show the relativistic effect, and simplify the calculation. 

In Fig. \ref{SIF}, we illustrate how this formula (\ref{spinHall}) provides an alternative calculation of the IF shift for relativistic electrons reflected from a potential step, considering the conservation of TAM and the angles of incidence and reflection. Notably, the results obtained from Eq. (\ref{spinHall}) are consistent with those from Eq. (\ref{Sif}), as demonstrated in Fig. \ref{SIF}. Therefore, the transverse IF shift is a direct manifestation of the spin Hall effect. The fact that the IF shift depends on the polarization of the incident particle suggests a strong connection between the transverse IF shift and coupled spin and orbit angular momentum involved in the spin properties of relativistic Dirac electrons.

Typically, the spin Hall effect is a phenomenon in which a spin-polarized current separates into two transverse components with opposite spin orientations. This effect arises in materials with spin-orbit coupling, where the spin of the charge carriers becomes coupled with their motion. The connection we observe between the transverse displacement and spin properties is reminiscent of the spin Hall effect. The transverse spin separation is driven by the intrinsic spin-orbit coupling involved in the Dirac equation, where the spin-to-orbital angular momentum conversion plays a crucial role in this process \cite{Bliokhpra17}.
In addition, the side jump effect refers to the lateral displacement experienced by a charge carrier when it undergoes scattering at an interface or in a non-uniform magnetic field \cite{skewscatteringprl,vignale2010ten}. It is a consequence of the Lorentz force acting on the charge carriers due to the presence of an electric field or a magnetic field gradient. The transverse IF shift shares the similarity with the side jump effect in terms of 
the pesudo spin-orbit coupling (see the term $\propto p_y \Sigma_z$ in Appendix \ref{appendixA}), which can be viewed as creating an effective magnetic field, and in turn related to the force.

\section{Discussion and Conclusion}

Summarizing, we have investigated the  GH and IF shifts of relativistic electrons interacting with a step potential. 
Starting with the  GH shift in Sec. \ref{GH}, which arises from the longitudinal displacement of the beam upon reflection, we have derived its expression using the Dirac current conservation and have showed its direct relation to the current within the evanescent wave in the transmission region. The GH shift diverges as the incidence angle approaches the critical angle, indicating its sensitivity to the proximity of total reflection and beam distortion. Importantly, we have observed that the GH shift remains the same for both spin-up and spin-down polarization along the $z$ direction.

In Sec. \ref{IF}, we have focused on the transverse IF shift, representing the displacement of the reflected beam in the $yox$ plane. Unlike the GH shift, the IF shift is much smaller in magnitude and reaches its maximum value for an incidence angle close to the critical angle during total reflection. The IF shift can also occur and become significant in presence of partial reflection. We have clarified that the IF shift arises from the current within the transmission region along the $x$ direction.
Furthermore, we demonstrated the equivalence of the IF shift to the spin Hall effect, based on  spin-to-orbital angular momentum conservation and relativistic effect. 
Remarkably, the IF shift displays a dependence on the spin polarization of the incident beam. In particular, opposite polarizations lead to shifts of equal magnitude but in opposite directions. In contrast, the side-jump mechanism describes the displacement of electrons during scattering, which is contingent on their spin orientation.
As a result, the behavior of the IF shift implies a resemblance to the transverse shifts observed in the side jump effect, hinting at potential connections between these phenomena.

However, there are several important avenues for future exploration. Our current analysis primarily focused on realistic electrons described by the Dirac equation, with eigenstates of spin polarization or helicity. However, considering arbitrary polarization (helicity) states would be valuable, as it allows us to explore the effects of different polarization states on the observed shifts. By incorporating this additional freedom \cite{Wang_2011}, we can achieve a more comprehensive understanding of the scattering behavior of relativistic electron beams and how their polarization properties influence the GH and IF shifts. Another noteworthy consideration is that, thus far, our assumptions have centered on plane waves in the spectrum having uniform polarization. Furthermore, investigating non-uniform polarization distributions, such as non-diffraction Bessel beams \cite{RobertPRX} and Bessel-Gaussian vortex beams \cite{Bliokhprl11,Barnettprl}, presents intriguing possibilities for future study.
Additionally, our analysis is conducted within the relativistic electron's scattering from a step potential. Future investigations can be dedicated to the effects of different potential profiles and incorporate additional factors, such as external fields, which could introduce supplementary contributions to the shifts. 

Last but not least, it is worth highlighting the recent experimental strides in electron vortex beams, which carry orbital angular momentum \cite{verbeeck2010production,uchida2010generation,science.1198804}. These achievements have not only offered diverse practical applications in microscopy, quantum information processing, and material characterization \cite{RevModPhys.89.035004}, but have also ignited a renewed interest in wave-packet dynamics of relativistic electrons \cite{Bliokhprl11,Bliokhprl12,Barnettprl,Zofiaprl17}, such as the realistic Hall effect and relativistic electron vortices. This experimental progress now serves as a platform for testing the these shifts. Hence, our findings emphasize the vital role of accounting for current conservation, interference effect, and spin-to-orbital angular momentum conversion in comprehending and manipulating these lateral shifts. Moreover, the polarization-dependent traits we observed present exciting prospects for innovating novel devices and applications in the devices using Dirac and topological materials.

\begin{acknowledgments}
This work was financially supported by the Basque Government through Grant No. IT1470-22, the Project Grant No. PID2021-126273NB-I00 funded
by MCIN/AEI/10.13039/501100011033 and by “ERDF A
way of making Europe” and “ERDF Invest in your Future,”
Nanoscale NMR and complex systems (Grant No. PID2021-
126694NB-C21), EU FET Open Grant EPIQUS (No. 899368), the ELKARTEK Program by the Basque
Government under Grant KK-2022/00041, BRTA QUANTUM Hacia una especialización armonizada en tecnologías
cuálnticas en BRTA.  X.C. acknowledges ayudas para contratos Ramón y Cajal–2015-2020
(Grant No. RYC-2017-22482). 
\end{acknowledgments}

\appendix

\section{Expectation values of spin and angular momentum and transverse IF shift}

\label{appendixA}

In this appendix, our objective is to derive the IF shift from the conservation of TAM. To achieve this, we start by calculating the expectation values of spin and angular momentum for the incident and reflected Dirac particle beams. In general, the total angular momentum is expressed as $\vec{J} = \vec{L} + \vec{\Sigma}$, where $\vec{L}$ is the orbital angular momentum operator, and the spin operator $\vec{\Sigma}$ has the three components, 
\begin{equation}
\Sigma_i = \frac{1}{2}
\left(
\begin{matrix} 
\sigma_i & 0 \\ 0 & \sigma_i
\end{matrix}
\right),
\end{equation}
with $i=x,y,z$. 
In our scenario, the $z$ component of the total angular momentum, $\langle J_z \rangle = \langle L_z \rangle + \langle \Sigma_z \rangle $, is conserved, leading to $ \langle J_z^\textrm{in} \rangle = \langle J^\textrm{r}_z \rangle$.

To perform the calculations, we utilize the FW transformation, given by:
\begin{equation}
U = \cos\left(\frac{\alpha}{2}\right) \mathds{1} + \frac{\beta (\vec{\alpha} \cdot \vec{p})}{|p|} \sin\left(\frac{\alpha}{2}\right),
\end{equation}
where $\alpha = \arctan(p/m)$. This unitary transformation is employed to diagonalize the Dirac Hamiltonian and obtain a more convenient representation where the positive and negative energy states are separated \cite{greiner2000relativistic}. It simplifies the calculations and allows for a clearer separation of different physical phenomena in the Dirac equation.
Using this transformation, we can express the expectation value of $\Sigma_z$ for the incident particle as:
\begin{eqnarray}
\langle \Sigma_z^\textrm{in} \rangle = \langle \Psi_\textrm{in}| \Sigma_z | \Psi_\textrm{in} \rangle = \langle \Psi_\textrm{in}| U^{\dag} U \Sigma_z U^{\dag} U | \Psi_\textrm{in} \rangle.
\end{eqnarray}
By substituting the wave function (\ref{psizin}), we can calculate the expectation value of $\Sigma_z$ for the incident relativistic electron as:
\begin{eqnarray}
\label{spinin}
\langle  \Sigma_z^\textrm{in} \rangle =\frac{ \mathcal{N}}{2} \left(1- \Delta \right),
\end{eqnarray}
where $\Delta= (1- m/E) \sin^2 \theta$ is a spin-orbit interaction parameter involving the incidence angle $\theta$, and $\mathcal{N} = \langle \Psi_\textrm{in} | \Psi_\textrm{in} \rangle = \sec^2\left(\alpha/2\right) =2E/(E+m)$. 
In the non-relativistic limit, as $\alpha$ and $\Delta$ approach 0 due to $E \simeq m$, we have $\langle \Sigma_z^\textrm{in} \rangle = 1/2$, which corresponds to the exact eigenvalue of the initially spin-up state, $|\uparrow \rangle$, in the Schrödinger frame.
Similarly, we can apply the FW transformation to the operators:
\begin{eqnarray}
\label{xFW}
Ux^\textrm{in}U^{\dag} &=& x^\textrm{in}_0+\frac{p_y \Sigma_z}{ E(E+m)} ,
\\
U p_y U^{\dag} &=& p_y.
\end{eqnarray}
In Eq. (\ref{xFW}), we have used $p_x=0$ and retained only the terms contributing to
the expectation values with respect to our initial state
for brevity. More detailed calculation and formulas can be found in Ref. \cite{greiner2000relativistic}.
By doing so, we obtain $\langle L_z^\textrm{in} \rangle = \langle \Psi_\textrm{in}| x^\textrm{in} p_y| \Psi_\textrm{in} \rangle =\langle \Psi_\textrm{in}| U^{\dag} U x^\textrm{in} U^{\dag} U p_y U^{\dag} U |\Psi_\textrm{in} \rangle$, yielding
\begin{equation}
\label{AMin}
\langle L_z^\textrm{in} \rangle  = \frac{ \mathcal{N}}{2}  \Delta + \langle x^\textrm{in}_0 \rangle p_y .
\end{equation}
As a consequence, we find, by using Eqs. (\ref{spinin}) and (\ref{AMin}), that:
\begin{equation}
\langle J_z^\textrm{in} \rangle = \langle L_z^\textrm{in} \rangle + \langle \Sigma_z^\textrm{in} \rangle = \frac{ \mathcal{N}}{2} + \langle x^\textrm{in}_0 \rangle p_y.
\end{equation}
Here we highlight the presence of spin-orbit interaction ($\propto p_y \Sigma_z$) in the second term of Eq. (\ref{xFW}),  also known as spin-to-orbital angular momentum conservation \cite{Bliokhpra17}. It plays the significant role in the scattering behavior of relativistic electrons and contributes to the fascinating phenomena, such as IF shift and spin Hall effect, observed in our study.

For the reflected case, we also find that the expectation value of the $z$ component of the total angular momentum is given by:
\begin{equation}
\langle J_z^\textrm{r} \rangle = \langle L_z^\textrm{r} \rangle + \langle  \Sigma_z^\textrm{r} \rangle = \frac{ \mathcal{N}}{2} (|r|^2- |r'|^2) + \langle x^\textrm{r}_0 \rangle p_y,
\end{equation}
where the terms are respectively calculated as follows:
\begin{eqnarray}
\langle  \Sigma_z^\textrm{r} \rangle &=& \frac{\mathcal{N}}{2} \left(1- \Delta \right) (|r|^2- |r'|^2),
\\
\langle L_z^\textrm{r} \rangle &=& \frac{ \mathcal{N}}{2} \Delta (|r|^2- |r'|^2)+ \langle x^\textrm{r}_0 \rangle p_y.
\end{eqnarray}
By further applying $ \langle J_z^\textrm{in} \rangle = \langle J^\textrm{r}_z \rangle$, we finally end up with the transverse IF shift, expressed as:
\begin{equation}
\delta S_x = \langle x^\textrm{r}_0 \rangle - \langle x^\textrm{in}_0 \rangle = \frac{\mathcal{N}}{2 p_y} \left[1-(|r|^2- |r'|^2) \right].
\end{equation}
This observation bears a resemblance to the Hall effect of light \cite{Hallprl04}. In addition, the whole calculations are also similar to the ones for Weyl semimetals \cite{Yangprl15,Jianprb17} and topological systems \cite{Yaoprb19}. However, the relativistic effects, specifically the intrinsic spin-orbit interaction, lead to the spin-dependence of the orbital angular momentum (and other observable orbital characteristics). This indicates intriguing phenomena and different characteristics in the scattering behavior of relativistic electron, or other Dirac spin-1/2 particles.

\bibliography{apssamp}

\end{document}